\documentclass[12pt]{iopart}
\usepackage{graphicx}

\begin{document}
\title[Flow Fluctuations]{Flow fluctuations and long-range correlations:
elliptic flow and beyond}

\author{Matthew Luzum}

\address{
CEA, IPhT, Institut de physique th\'eorique de Saclay, F-91191
Gif-sur-Yvette, France}

\ead{matthew.luzum@cea.fr}

\begin{abstract}
These proceedings consist of a brief overview of the current understanding of collective behavior in relativistic heavy-ion collisions.  In particular, recent progress in understanding the implications of event-by-event fluctuations have solved important puzzles in existing data --- the ``ridge'' and ``shoulder'' phenomena of long-range two-particle correlations --- and have created an exciting opportunity to tightly constrain theoretical models with many new observables.
\end{abstract}

\section{Introduction}
It is well established that the medium created in relativistic heavy-ion collision is characterized by strong collective behavior.  In particular the large ``elliptic flow'' anisotropy in multiparticle correlations has been interpreted as evidence of a low-viscosity, nearly-perfect fluid being created in these experiments.

Even so, it was only recently that we have begun to appreciate the full impact of this collective evolution on these correlation measurements, due to the non-trivial effect of event-by-event fluctuations.  In these proceedings I review this recent progress and speculate on future prospects for new flow measurements to 
deepen our understanding of all stages of the evolution of heavy-ion collisions.
\section{Two-particle correlations}
\begin{figure}
\includegraphics[width = .57\linewidth]{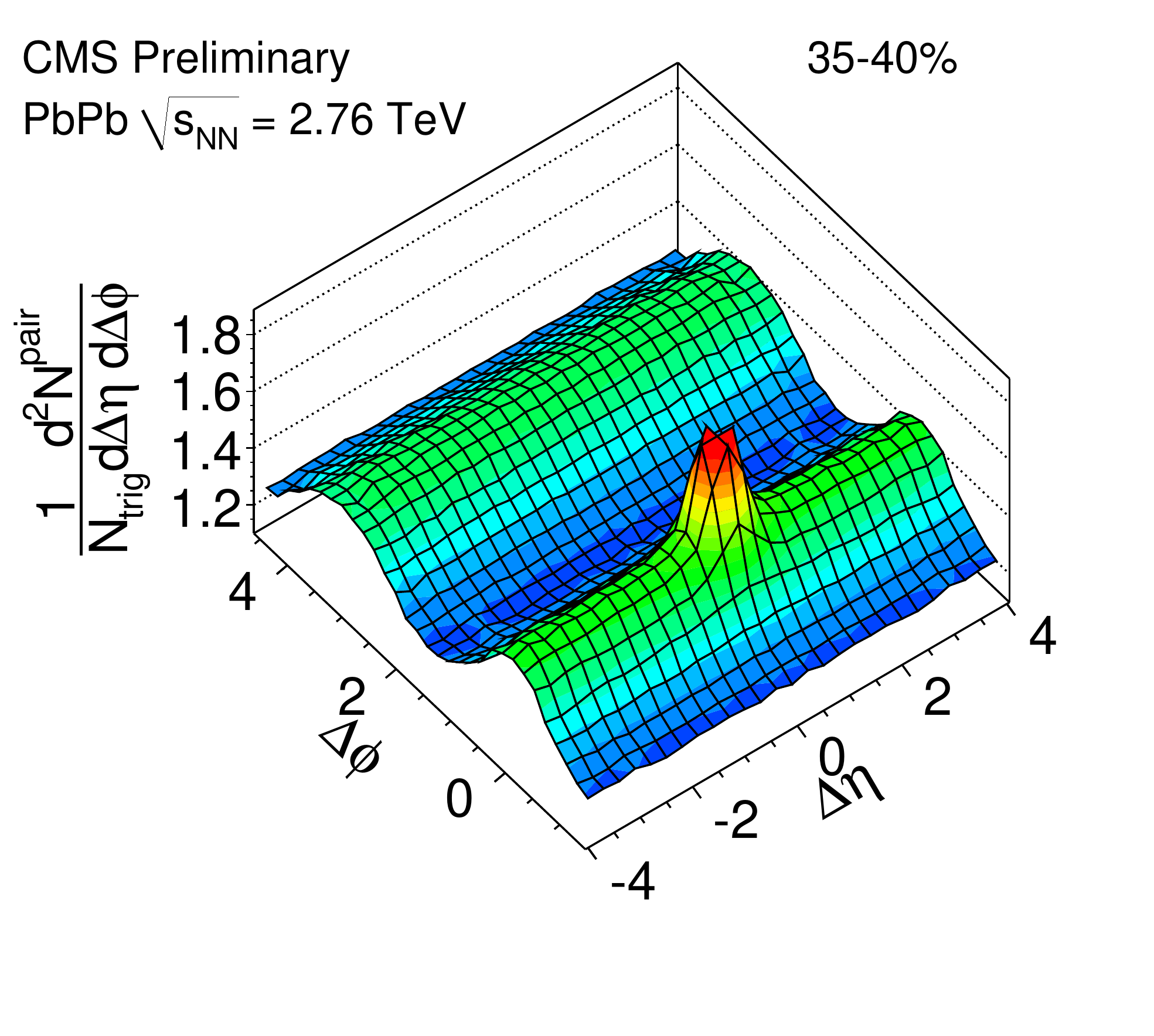}
\includegraphics[width = .43\linewidth]{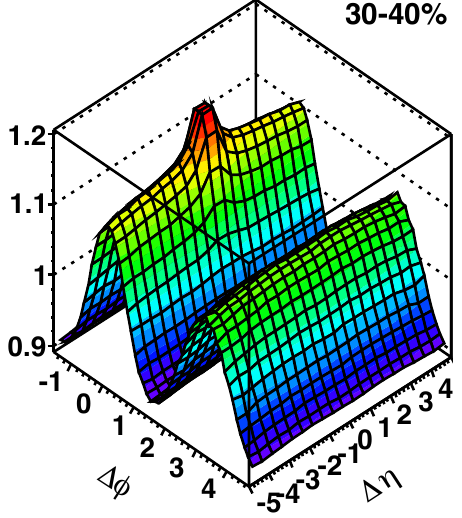}
\caption{\label{corr}Preliminary two-particle correlation data for charged hadrons in Pb-Pb collisions from the CMS (left) \cite{CMS} and ATLAS (right) \cite{ATLAS} collaborations as presented at this Quark Matter 2011 conference.  The CMS analysis restricts one particle of the pair to have transverse momentum $2 < p_t < 4$ GeV and the other $4 < p_t < 6$ GeV while ATLAS uses all particles with $2 < p_t < 3$ GeV.}
\end{figure}

Correlations between particles detected in heavy-ion collisions have been a powerful tool in understanding the properties and dynamics of the collision system.  The simplest and most ubiquitous is a two-particle correlation, where the distribution of pairs of particles from the same event are calculated and averaged over events.  Since the azimuthal orientation of each collision is uncontrolled, one can only measure rotationally symmetric quantities.  Therefore, one must fix the relative azimuthal angle of the particle pair $\Delta\phi = \phi^{(1)}-\phi^{(2)}$ and measure the average $dN_{\rm pairs}/d\Delta\phi$ (the pseudorapidity dependence is not so constrained, but is also often presented as a function of $\Delta\eta$).  Two recent examples are shown in Fig.~\ref{corr}.  

At small relative pseudorapidity are features that are familiar from smaller collision systems like proton-proton --- e.g., the positive correlations near $\Delta\phi=\Delta\eta=0$ due to Bose-Einstein correlations and jets.  However, a significant portion of the measured correlation is unique to heavy-ion collisions and has been interpreted as being generated from strong collective motion of the system, or ``flow'' --- in particular the large second Fourier component in $\Delta\phi$ that is present even for particles separated by a very large relative pseudorapidity $\Delta\eta$.

The standard picture of this flow correlation is as follows.  One imagines that particles in a given collision event are emitted at the end of collective evolution of the system according to a probability distribution (which depends on the initial geometry and subsequent collective expansion as described below).  The azimuthal dependence of this final particle distribution can be generally written as a Fourier series with respect to the azimuthal angle $\phi$ of the momentum of the outgoing particle:
\begin{equation}
\label{dist}
\frac {2\pi} {N}\frac {dN} {d\phi}  
=  1 + \sum_{n=1}^\infty 2 v_n \cos n(\phi-\Psi_{n}) 
=  \sum_{n=-\infty}^\infty v_n e^{in\Psi_n} e^{-in\phi} .
\end{equation}
Here $v_n$ are the flow coefficients, while $\Psi_n$ is the reference angle for each harmonic $n$ that is defined as the phase of the complex Fourier coefficient $\langle e^{in\phi} \rangle = v_n e^{in\Psi_n}$, or equivalently the angle defined by $\langle \sin n(\phi-\Psi_n)\rangle=0$.  Note that, when defined this way, both $\Psi_n$ and $v_n$ can in principle depend on transverse momentum and pseudorapidity.

If there is strong collective behavior, each particle is emitted independently and the pair distribution is given simply by the product of single-particle distributions.  For example, a single Fourier component of the pair distribution is given by
\begin{equation}
\label{vndelta}
\left\langle \langle e^{in\left(\phi^{(1)}-\phi^{(2)}\right)} \rangle \right\rangle 
\stackrel{\rm{(flow)}}{=} 
\left\langle \langle e^{in\phi^{(1)}}\rangle \langle e^{-in\phi^{(2)}}\rangle \right\rangle 
= \left\langle v_n^{(1)}v_n^{(2)}e^{in(\Psi_n^{(1)}-\Psi_n^{(2)})}\right\rangle ,
\end{equation}
%
%
%
%
%
%
where the small inner brackets represent an average over pairs in a single event and the large outer brackets indicate an average over events in a centrality class.  The entire azimuthal dependence can then be written as
\begin{equation}
\label{pair}
\frac {2\pi} {N_{\rm pairs}}\left\langle\frac {dN_{\rm pairs}} {d\Delta\phi}\right\rangle 
\stackrel{\rm{(flow)}}{=} 
1 +  \sum_{n=1}^\infty 2 \left\langle v_n^{(1)}v_n^{(2)} \right\rangle \cos n(\Delta\phi).
\end{equation}
This form is typically called a ``flow'' correlation, while any intrinsic pair correlations in a given event break this single-event factorization, and such an extra contribution to the observable can be termed ``non-flow''.
It should be noted that, in addition to independent emission, this commonly-written form of Eq.~(\ref{pair}) also contains the implicit assumption that the orientation angle $\Psi_n$ is a global quantity that depends little on transverse momentum or pseudorapidity (or more precisely, $\Psi_n^{(1)}\simeq\Psi_n^{(2)}$ for all pairs in the analysis), as well as the assumption that multiplicity fluctuations in the centrality bin are small enough that one can take the $N_{\rm pairs}$ factor out of the average.

Until recently it was assumed that, due to the symmetry of colliding identical spherical nuclei, the following are true:  $\Psi_n \simeq 0$ when defined with respect to the impact parameter,  the flow coefficients for $n$=odd are odd in (pseudo)rapidity, and therefore only even terms $n=2,4$ are important near mid-rapidity. Thus, the sizeable second Fourier harmonic, largely independent of relative pseudorapidity,  that is seen in these two-particle correlations has been interpreted as coming from the ``elliptic flow'' response to an initial almond-shaped collision overlap region, and indeed serves as a measurement of the coefficient $v_2$.  One can see this by eye in Fig.~\ref{corr} as the two large elongated bumps centered at $\Delta\phi=0$ and $\Delta\phi=\pi$.  Upon closer inspection, however, it becomes apparent that the peaks are not identical.
 Thus, the $\Delta\phi$ dependence can not be entirely described by only a second Fourier component, even when gaps are imposed in the relative pseudorapidity of the particle pairs to suppress most known non-flow correlations.  

The remaining long-range (and until recently, presumed to be non-flow) contribution has been a topic of significant study.   
To do this, one must choose a model scheme to separate the flow from non-flow.  A common example is the Zero-Yield-At-Minimum (ZYAM) prescription~\cite{Adler:2005ee}, which posits a small non-flow correlation coming from a set of particles that have zero production at one or more values of $\Delta\phi$.  Next, one must also posit the value of each flow coefficient $v_n$, though in most cases only a non-zero $v_2$ is used.  After this flow-subtraction procedure, there typically remains a narrow ``ridge'' feature at $\Delta\phi=0$, as well as a broad feature on the away side, which often displays a ``shoulder'' or ``cone'' structure with a dip at $\Delta\phi=\pi$.  Similar features can also now be seen in extremely central collisions without subtraction~\cite{ATLAS,ALICE}.

Any event-by-event fluctuation, however, will break the apparent symmetry of the collision system. 
Flow fluctuations have been known to be important for several years, but initially only the effect on elliptic flow $v_2$ was studied~\cite{Alver:2006wh}.  A crucial step forward was the realization that event-by-event fluctuations imply that other coefficients, especially $v_3$, should be non-negligible~\cite{Alver:2010gr}.  Thus, the previous flow-subtraction procedures did not actually remove the flow correlation, and in fact, flow correlations alone can potentially generate the entire observed two-particle correlation at large $\Delta\eta$~\cite{Luzum:2010sp}.  There is a growing consensus that this is indeed the case, but to test this one must first specify all the properties we expect such flow correlations to have, and then search the data for evidence of a non-flow contribution.
\section{Hydrodynamics}
The state-of-the art description of the bulk evolution of heavy-ion collisions is the hydrodynamic framework~\cite{SchenkeQM}.  Such models characterize the collision medium as an approximately thermalized relativistic fluid with a few macroscopic properties such as viscosity and equilibrium equation of state, subject to initial conditions and a prescription for particles to ``freeze-out'' from the fluid at the end of the evolution.  The generic picture is that the spatial anisotropy in the density of the system at the onset of collective evolution is converted into a momentum anisotropy that manifests itself in correlations among the final particles.  Event-by-event fluctuations in the final distribution~(\ref{dist}), then, ultimately stem from fluctuations in the initial geometry, which are evolved forward in time in a hydrodynamic framework.

The early-time, pre-equilibrium dynamics are not yet understood in detail.  Typically a few simple models are used to probe the range of results that can reasonably be expected.  However, there are generic features that are expected.  For example, most models predict flux-tubes or string-like structures to appear at early times that are extended longitudinally.  These structures ultimately cause very long-range correlations, like those seen in data (see Fig.~\ref{corr}).  The size of event-by-event fluctuations, and thus harmonics like $v_3$ that are generated by fluctuations, should also vary with centrality in a predictable way due to the change in size of the system.  In addition, the emission of approximately thermalized matter at the end of hydrodynamic evolution generically results in a mass ordering of each flow coefficient, like that seen experimentally for $v_2$, with smaller values at a fixed $p_t$ for heavier particles.

The hydrodynamic response to the initial geometry
has been extensively studied for smooth and symmetric initial conditions, and more recently with event-by-event fluctuating initial conditions~\cite{Holopainen:2010gz,Qin:2010pf,Qiu:2011iv,Gardim:2011qn,Werner:2010aa,Schenke:2010rr} in addition to transport calculations~\cite{Xu:2010du,Ma:2010dv}.  
Thus, much is already known.  However, work is ongoing to more precisely characterize this (non-linear) hydrodynamic response in order to more precisely extract properties of the collision system.  To that end, I describe a systematic approach, inspired by Ref.~\cite{Teaney:2010vd}, which more clearly explains the reason behind simple approximate hydrodynamic response relations that have been found such as $v_2\propto\varepsilon_2$, and suggests how to systematically implement corrections.
%
%

Take the 2D Fourier transform of the initial transverse density $\rho(\bf{x})$ and expand in harmonics and in powers of $k\equiv\sqrt{{\bf k}^2}$:
\begin{equation}
\frac {\rho({\bf k})}{\rho(0)} 
\equiv \frac {\int d^2x \rho(\bf{x}) e^{i \bf{k}\cdot \bf{x}}} {\int d^2x \rho(\bf{x})}
= \sum_{n=-\infty}^{\infty} \sum_{m=0}^{\infty}  \frac {i^m} {m!}  \varepsilon_{m,n} k^m e^{-i n\phi_k } ,
\end{equation}
with
\begin{equation}
\varepsilon_{m,n} = \frac {m!}{2^m(\frac{m+n}{2})!(\frac{m-n}{2})!}  \{ r^m e^{in\phi}\} ,
\end{equation}
and
\begin{equation}
\{ \ldots \} = \frac{\int  d^2x \rho({\bf{x}}) \ldots} {\int  d^2x \rho({\bf{x}})},
\end{equation}
where now $\phi$ is a spatial coordinate.
Note that the moments $\varepsilon_{m,n}$ are only non-zero for $m\geq |n|$ and $(m-n) =$ even.  Equivalently, and perhaps more naturally, one can do the same decomposition with the log of the Fourier transform to obtain the cumulants $W_{n,m}$~\cite{Teaney:2010vd}, but the higher order expressions are more complicated to write.

The final particle distribution at a given momentum can then be regarded as simply a function of (an infinite number of) moments of the initial density distribution. 
This is useful for two reasons.  First, hydrodynamics is a long-wavelength description of a system that is only valid if short-distance dynamics is not important.  Thus, we expect the hydrodynamic response to be insensitive to short-distance features represented by moments with high powers of $k$ --- i.e., viscous effects dampen modes with larger $m$.  We can then expect to be able to truncate this series and write a hydrodynamic response function of a finite number of moments, and if the anisotropy is not large compared to the system size, we can also write it as a Taylor series with a finite number of terms.   Second, it is a nice way of clearly specifying the symmetry constraints --- each flow coefficient $v_n$ can only depend on aspects of the initial density with the correct symmetry properties.

Take as an example elliptic flow.  The second complex Fourier coefficient $\langle e^{i2\phi} \rangle = v_2 e^{i2\Psi_2}$ can only depend on combinations of moments that are dimensionless and have the correct symmetries --- e.g., a rotation of the system $\phi \to \phi + \delta$ results in a multiplicative factor $e^{i 2 \delta}$, and if there is a reflection symmetry about some plane in the initial state, it must also be present.   Choosing a coordinate system such that $\{x\}=\{y\}=0$, the lowest order moments are $\varepsilon_{2,0}(\propto\{r^2\})$ and $\varepsilon_{2,2}(\propto\{r^2e^{2i\phi}\})$.  Thus, to lowest order in $m$, $v_2$ can only depend on the dimensionless ratio  
\begin{equation}
\varepsilon_{PP}e^{2i\Phi_{PP}} \equiv -\frac{\{r^2e^{2i\phi}\}}{\{r^2\}},
\end{equation}
where $\varepsilon_{PP}$ and $\Phi_{PP}$ are the standard participant eccentricity and participant plane, respectively.  Since typically $\varepsilon_{PP}\ll1$, we can write the function as a power series in $\varepsilon_{PP}$.  Thus, to first order we have
\begin{equation}
\langle e^{i2\phi} \rangle 
\equiv v_2 e^{i2\Psi_2} 
= C\ \varepsilon_{PP}e^{2i\Phi_{PP}}
\equiv - C\ \frac {\{r^2 e^{2i\phi}\}} {\{r^2\}} ,
\end{equation}
for some (real) coefficient $C$ that encodes information about, e.g., viscosity, equation of state, etc.  This is the familiar relationship $v_2\propto\varepsilon_2$, which has long been known for hydrodynamic simulations with smooth initial conditions, but also works to a reasonable approximation in lumpy, event-by-event hydrodynamic calculations~\cite{Holopainen:2010gz,Qin:2010pf,Qiu:2011iv}.  In this context, one can see that the expression is just the first term in a controlled expansion, with corrections coming from terms higher order in $m$ ($\varepsilon_{4,2}\propto\{r^4e^{2i\phi}\}$), or in the Taylor series ($\varepsilon_{PP}^3$).   

As an aside, it should by now be clear that $v_2$ can not depend on a term that is linear in $\varepsilon_{3,3}$, as proposed in Ref.~\cite{Qin:2010pf}, because it does not have the correct symmetries.  It would have to depend on combinations like $\varepsilon_{3,3}\varepsilon_{3,1}^*$ or $\varepsilon_{3,3}^2 \varepsilon_{2,2}^{*2}$, etc.

Similar approximate proportionality relations have been found to reasonably well describe the results for $v_3$~\cite{Qin:2010pf,Qiu:2011iv} and $v_1$~\cite{Gardim:2011qn}, while $v_4$ and $v_5$ are more complicated~\cite{Qiu:2011iv}.  In retrospect, this is unsurprising since the possible $v_4$ terms $\varepsilon_{4,4}$ and $\varepsilon_{2,2}^2$ are typically of the same size, with the former being more important in central collisions and the latter more important in peripheral collisions, in agreement with results from Ref.~\cite{Qiu:2011iv}.  Explicitly this could read something like:
\begin{equation}
\langle e^{i4\phi} \rangle 
= v_4 e^{in\Psi_4} 
= C_1 \frac {\{r^2 e^{2i\phi}\}^2} {\{r^2\}^2} + C_2 \frac {\{r^4 e^{4i\phi}\}} {\{r^4\}} 
\end{equation}
A similar statement can be made about the dependence of $v_5$ on $\varepsilon_{5,5}$ and $\varepsilon_{3,3}\varepsilon_{2,2}$.

The hydrodynamic response has been confirmed to significantly damp higher harmonics~\cite{Alver:2010dn}, in agreement with data~\cite{ATLAS}.  Thus, once the hydrodynamic response is mapped out for the first $\sim$6 flow harmonics to the order desired, for each centrality and each set of parameter values, all useful information about the hydrodynamic model is known.  This makes it clear exactly what properties of the initial geometry are important, and allows one to quickly calculate correlations arising from an arbitrary set of initial conditions.
\begin{figure}
\includegraphics[width=0.3\linewidth]{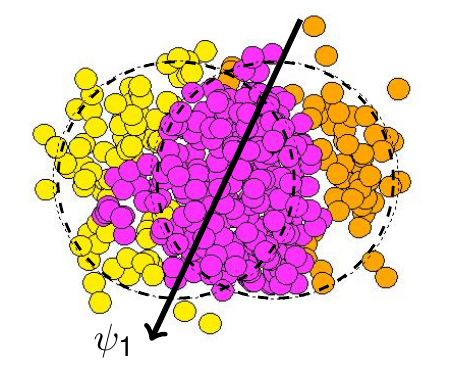}
\includegraphics[width=0.35\linewidth]{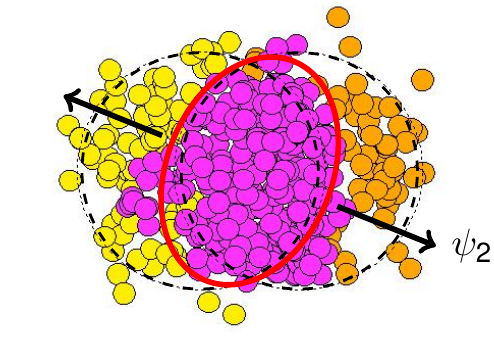}
\includegraphics[width=0.35\linewidth]{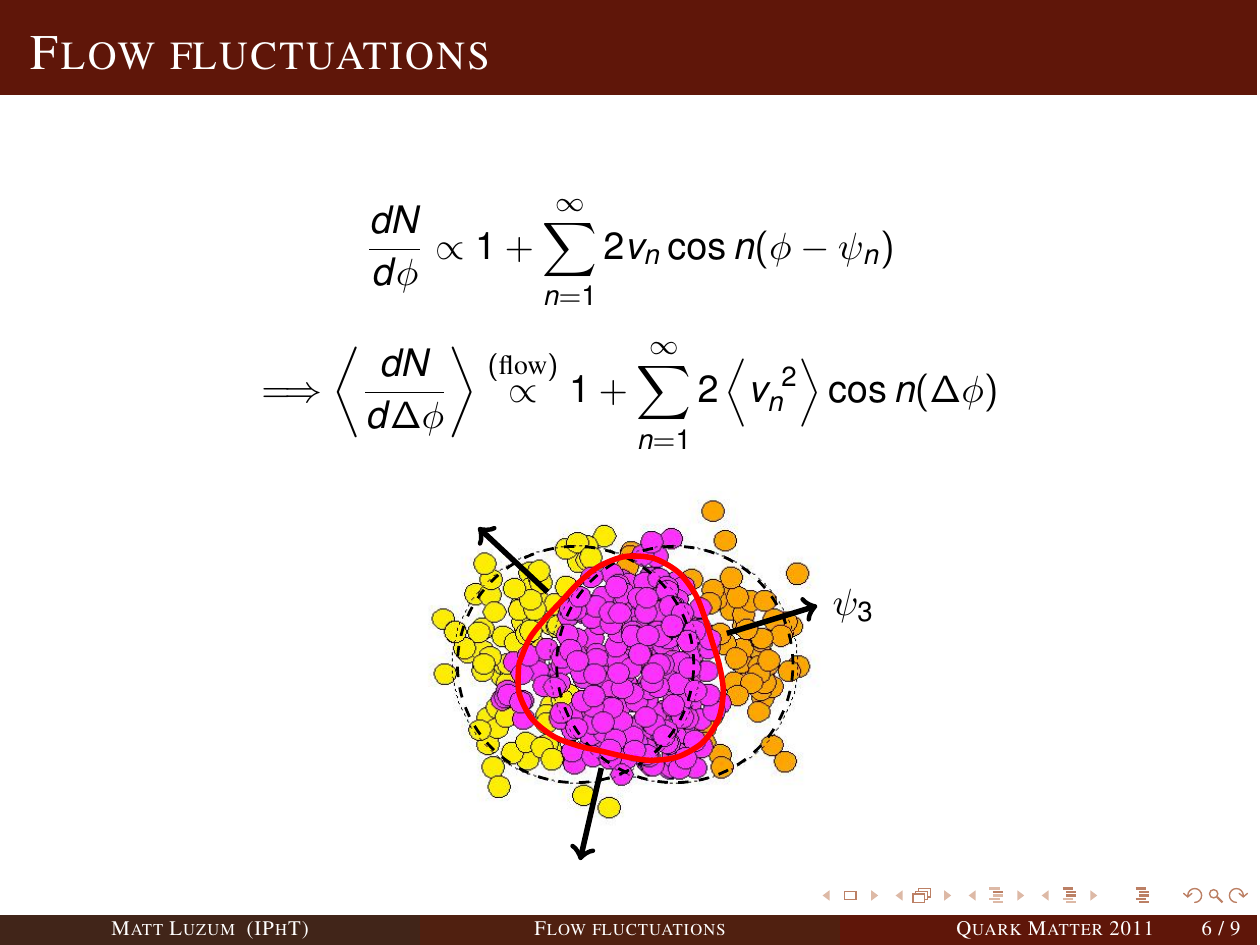}
\caption{Schematic diagram illustrating the simultaneous orientation of directed ($\Psi_1$), elliptic ($\Psi_2$), and triangular flow ($\Psi_3$) in relation to the initial distribution of participant nucleons in a single event from a Glauber Monte Carlo~\cite{Alver:2008aq}.}
\end{figure}
\section{Flow vs. data}
Now that we have a picture of flow, one can look in detail at the long-range two-particle correlation data to see whether they quantitatively agree with this picture, or if one should instead conclude that other correlations are likely to be present.

Already mentioned are a number of generic qualitative features that are in agreement with this flow picture.  The observed correlation extends to large values of $\Delta\eta$, with all Fourier components having little dependence on relative pseudorapidity, in agreement with flow resulting from the longitudinally-extended structure of most models of the initial state.  (Note that the flow picture does \textit{not} predict or require exactly zero dependence on $\Delta\eta$).   In addition is the lack of non-zero high Fourier components, consistent with the observation of strong damping of higher modes in hydrodynamics (especially with a non-zero viscosity).   Finally, the mass ordering predicted by hydrodynamics and seen for $v_2$ is also present in the other Fourier harmonics~\cite{ALICE}.

More convincing, of course, is the quantitative agreement with predictions.  The centrality dependence and size of the second and third harmonics agree with that predicted for $v_2$ and $v_3$~\cite{Alver:2010gr, Alver:2010dn}.  The $p_t$-dependence  of all the harmonics as well as how they vary with the orientation of the particle pair with respect to the measured event plane $\Psi_2$ is also exactly as expected from flow correlations~\cite{Luzum:2010sp}, with the exception of the first harmonic, which non-trivially displays just the $p_t$-dependence expected from a combination of flow plus the well-known momentum-conservation correlation that naturally arises when a finite number of final-state particles share a fixed total transverse momentum and only contributes to the first Fourier harmonic~\cite{Luzum:2010fb} (note that having zero transverse momentum does not mean $v_1=0$, only that $(p_t\cdot v_1)$ integrates to zero).

Finally, this Quark Matter conference has seen an even more stringent test --- the factorization of each Fourier harmonic into a function of the transverse momentum of the first particle times the $p_t$ of the second;  i.e., an ansatz something like
\begin{equation}
\left\langle v_n^{(1)}v_n^{(2)} \right\rangle = v_n^{(1)}\{2\}(p_t)\times v_n^{(2)}\{2\}(p_t) .
\end{equation}
Note that this form is even more restrictive than the independent emission assumption of a flow correlation, which implies only single-event factorization.  For this to be true after averaging over events requires that the event-by-event fluctuation can be represented by a global factor for each harmonic in each event --- e.g., the $p_t$ dependence is similar in every event and only scales by a constant factor event-by-event.  Although this feature is not generically required in the flow picture, and hasn't yet been thoroughly tested in hydrodynamic calculations, it is a reasonable expectation and a very non-trivial test of the existence of non-flow correlations like an away-side jet correlation.  Remarkably, this factorization has been found to hold quite well~\cite{AdareQM} for the bulk of particles and for every harmonic except the first, which is contaminated by the momentum-conservation correlation mentioned above.  One only sees the factorization begin to break down for particles above $\sim$4--5 GeV, and for very peripheral collisions, as expected.

Hence the growing consensus that, aside from the exceptions listed, flow correlations from a hydrodynamic response of fluctuating initial geometries are responsible for the entire two-particle correlation structure at large $\Delta\eta$.

A comment should be made here about the so-called ``event plane'' flow analysis, which is also often used to extract flow coefficients $v_n$.  Exactly what is measured depends on a quantity called the ``event plane resolution'', which is essentially the square root of the number of particles used in the analysis times the magnitude of the flow coefficient itself $v_n\sqrt{N}$~\cite{Ollitrault:2009ie}.  In most cases, the analysis is sensitive only to two-particle correlations.
If the event plane resolution parameter is very large (as for some of the new measurements of elliptic flow at the LHC), the result begins to be sensitive to higher particle correlations.  The difference between the result of an event-plane analysis and a two-particle correlation analysis in such a case (like the difference between two-particle and multiparticle analyses) thus depends on non-flow correlations that may be present as well as fluctuations, but only in a particular combination that cannot be disentangled~\cite{Ollitrault:2009ie}.  One must therefore use caution when comparing $v_n$ from an event plane flow measurement to the same quantity extracted from a two-particle correlation, in an effort to make a statement about non-flow correlations.   If the two values agree, it does not indicate an absence of non-flow correlations, only that the event plane resolution is not large.  If the values do not agree, it could be due entirely to flow fluctuations, and similarly this does not indicate the existence of non-flow correlations.  In neither case does it address the question of this section, whether long-range correlations are due only to flow --- in other words, taking $v_n$ from a dedicated flow analysis, plugging them into Eq.~(\ref{pair}), and comparing to the total two-particle correlation does not necessarily test the presence of non-flow.   Likewise, this means that there exists no independent measurement of the flow coefficients that one can use in subtraction schemes such as ZYAM, and this is why assumptions are always necessary.  A general and rigorous way to separate flow from non-flow correlations does not exist.
\section{New flow observables}
With the knowledge that long-range two-particle correlations are likely dominated by flow correlations comes the opportunity for precise measurement of many independent flow observables with little non-flow contamination.  The new measurements of $v_3$, $v_4$, $v_5$, and $v_6$ presented at Quark Matter 2011~\cite{PHENIX,ALICEflow,ATLAS,CMSflow,STAR} will undoubtedly guide theoretical progress for some time, and a dedicated analysis of a rapidity-even component of $v_1$~\cite{Luzum:2010fb,Teaney:2010vd} will exhaust the possibilities of strictly two-particle correlations.  Already from simultaneous measurements of $v_2$ and $v_3$, one can rule out models for the initial conditions (e.g., a particular MC-KLN implementation seems to have insufficient relative eccentricity fluctuations)~\cite{PHENIX,Alver:2010dn}.

However, an even larger set of observables opens up with correlations between three or more particles.  Some are familiar, like the fourth cumulant elliptic flow $v_2\{4\}$, which can be generalized to all harmonics, like the recently measured $v_3\{4\}$~\cite{ALICEflow}.  Because of the existence of fluctuations, this measures independent information about the event-by-event distribution of $v_n$.  However, many other observables are possible, giving information about, e.g., correlations between the various orientation angles $\Psi_n$~\cite{Bhalerao:2011yg}, and providing a wealth of independent information.  For example, the comparison of $v_2\{4\}$ compared to a two-particle correlation measurement $v_2\{2\}$ had already indicated the above conclusion about the size of fluctuation in the same MC-KLN model~\cite{Bhalerao:2011yg}.  

With this information from many independent measurements, it should be possible not only to more precisely characterized properties of the thermalized medium, such as viscosity, but to significantly constrain the nature of fluctuations and the properties of the pre-equilibrium system in general.
\section{Summary}
In these proceedings, I have reviewed the emerging realization of the importance of event-by-event flow fluctuations in heavy ion collisions.  With this knowledge, long-range two-particle correlations (e.g., ``Mach cone'' and ``ridge'' structures) are naturally explained as a result of a collective response to an initial density distribution that fluctuates event-by-event.  This also implies many new flow observables that can be precisely measured and used to significantly restrict theoretical models --- not only other Fourier components of two-particle correlations like triangular flow, but a wealth of correlations between three or more particles --- and which will provide exciting opportunities to improve our knowledge of the entire collision evolution.
\ack
I thank Jean-Yves Ollitrault for useful input. This work is funded by ``Agence Nationale de la Recherche'' under grant
ANR-08-BLAN-0093-01.
\section*{References}
\bibliography{QMproceedings}{}
\end{document}